\documentclass[draft,twoside]{article}

\usepackage{amsmath,amssymb, amsthm,amsfonts,amscd,mathrsfs,pifont} 
\usepackage{eucal}  % Euler fonts
\usepackage{verbatim}     % I need the verbatim package here.
\usepackage[dvips]{graphicx}
\usepackage{hyperref}

\newtheorem{thm}{Theorem}[section]

\newtheorem{lemma}[thm]{Lemma}

\newtheorem{thm*}{Theorem}[]
\newtheorem{cor*}[thm*]{Corollary}
\newtheorem{claim*}[thm*]{Claim}
\newtheorem{lemma*}[thm*]{Lemma}
\newtheorem{prop*}[thm*]{Proposition}
\newtheorem{conj*}[thm*]{Conjecture}

\theoremstyle{definition}

\newtheorem{question*}{Question}[section]

\newtheorem{defn*}{Definition}

\theoremstyle{remark}

\newcommand{\BB}[1]{\ensuremath{\mathbb{#1}}}

\newcommand{\R}{\ensuremath{\BB{R}}}

\newcommand{\C}{\ensuremath{\BB{C}}}

\newcommand{\bs}{\ensuremath{\boldsymbol}}
\newcommand{\mf}{\ensuremath{\mathfrak}}
\newcommand{\wt}{\ensuremath{\widetilde}}

\newcommand{\la}{\ensuremath{\langle}}
\newcommand{\ra}{\ensuremath{\rangle}}

\newcommand{\transpose}{\ensuremath{\mathsf{T}}}

\newcommand{\ip}[1]{\mathrm{Im}(#1)}

\newcommand{\ul}[1]{\underline{#1}}

\DeclareMathOperator{\sgn}{sgn}

\DeclareMathOperator{\erfc}{erfc}

\DeclareMathOperator{\Pf}{Pf}

\newcommand{\mytitle}{Correlation Functions for $\beta=1$ Ensembles of
  Matrices of Odd Size} 

\numberwithin{equation}{section} 

\numberwithin{equation}{section}

\begin{document}
\title{\bf \mytitle}  
\author{\sc Christopher D.~Sinclair\footnote{This research was
  supported in part by the National Science Foundation (DMS-0801243)}}
\maketitle

\begin{abstract}
  Using the method of Tracy and Widom we rederive the correlation
  functions for $\beta=1$ Hermitian and real asymmetric ensembles
  of $N \times N$ matrices with $N$ odd.
\end{abstract}

\section{Introduction}

The starting point for many results concerning the spectral
theory of random matrices is the derivation of a determinental or
Pfaffian form for the correlation functions of the eigenvalues.
Perhaps the most well-studied ensembles are Hermitian ensembles having
joint probability density function (JPDF) of the form 
\begin{equation}
\label{eq:13}
\Omega(\bs \lambda) = \frac{1}{Z} \bigg\{ \prod_{n=1}^N
w(\lambda_n) \bigg\} | \Delta(\bs \lambda) |^{\beta} 
\end{equation}
where $Z$ a normalizing constant, $w: \R \rightarrow [0,\infty)$ is a
{\em weight} function and $\Delta(\bs \lambda)$ is the $N \times N$
Vandermonde determinant in the coordinates of $\bs \lambda$.  The
parameter $\beta$ is called the {\em inverse temperature} parameter
due to its physical interpretation in the study statistics of
log-gases, another area where joint densities of the form
(\ref{eq:13}) arise.  When $w$ is Gaussian and $\beta=1,2$ or $4$,
$\Omega$ is the JPDF of the ensemble of $N \times N$ real symmetric
($\beta=1$), complex Hermitian ($\beta=2$) and self-dual ($\beta=4$)
matrices, the independent entries of which are chosen with Gaussian
density.  The details of this derivation are given in \cite{MR2129906}
and the interpretation of $\Omega$ in the study of
log-gases is given in \cite{forrester-book}.  We will denote the
ensemble with joint density $\Omega$ by $w \beta$E.  

The $n$th correlation function of $w\beta$E is defined by 
\[
R_n(\bs \lambda) = \frac{1}{(N-n)!} \int_{\R^{N - n}}
\Omega(\bs \lambda \vee \bs \xi) \, d\mu^{N-n}(\bs \xi); \qquad n=1,2\ldots,N,
\]
where $\bs \lambda \vee \bs \xi = (\lambda_1, \ldots,
\lambda_n, \xi_1, \ldots, \xi_{N-n})$ and $\mu$ is Lebesgue measure on
$\R$.  That is, after 
renormalization, $R_n$ gives the $n$th marginal probability density of
$\Omega$. 

When $\beta=2$, Fubini's Theorem together with elementary row and
column operations on the Vandermonde determinant in the integrand lead
to the determinental formula
\[
R_n(\lambda) = \det\left[
K_N(\lambda_j, \lambda_k)
\right]_{j,k=1}^n,
\]
where $K_N$ is the kernel of a certain operator on $L^2(w \, d\mu)$.

Following pioneering work of Dyson \cite{MR0278668},
Mehta derived a Pfaffian\footnote{In fact Dyson, Mehta and many who
  followed, expressed the correlation functions in terms of {\em Quaternion}
    determinants, but in the more recent literature these are usually
  expressed in terms of Pfaffians.}  form for the correlation functions of
the Gaussian $\beta=1$ and $4$ Hermitian ensembles \cite{MR0277221}.
This was repeated for general weights by Mehta and Mahoux
\cite{MR1190440}, except for the case $\beta=1$ and $N$ odd.  This
last remaining case was given by Adler, Forrester 
and Nagao \cite{MR1762659}.  In each of these cases, the Pfaffian 
formulation of $R_n$ is given by  
\[
R_n(\lambda) = \Pf \left[
K_N(\lambda_j, \lambda_k)
\right]_{j,k=1}^n,
\]
where here, $K_N$ is a $2 \times 2$ matrix which is the kernel
of an operator on $L^2(w \, d\mu) \times L^2(w \, d\mu)$. (The matrix
$K_N$ depends on $\beta$ and has a slightly different
structure depending on whether $N$ is even or odd).  

Much of the interest in the spectral theory of random matrices
revolves around the eigenvalue statistics as $N \rightarrow \infty$.
It is for this reason that the determinental/Pfaffian formulations for the
correlation functions are so important: $N$ appears as a parameter in
the kernel and in many cases, $K_N$ be analyzed as $N \rightarrow
\infty$.  In contrast, the number of integrations necessary to arrive  
at the correlation functions via their definition, increases with
$N$---a situation which is not as easily handled.  

These determinental/Pfaffian forms of the correlation functions 
have since been derived by various different methods.  Of particular
note is the method of Tracy and Widom who, for all cases except
$\beta=1$, $N$ odd, derive the determinant/Pfaffian forms of the
correlation functions using the fact that, if
$\mathbf{A}$ is a $T \times N$ and $\mathbf B$ is an $N \times T$
matrix, then 
\begin{equation}
\label{eq:1}
\det(\mathbf I - \mathbf{AB}) = \det(\mathbf I - \mathbf{BA}),
\end{equation}
where on the left hand side of this equation $\mathbf I$ is the $T
\times T$ identity matrix, and on the right hand side it is the $N
\times N$ identity matrix.  
(Equation~(\ref{eq:1}) is more generally true; a fact which Tracy
and Widom exploit to great advantage in a number of other situations).
When $\beta=2$ this identity leads immediately to 
the determinental correlation functions whereas, when $\beta=1$ or
$\beta=4$, the path 
from (\ref{eq:1}) to the Pfaffian correlation functions is more convoluted.
The reason for this extra difficulty is that (\ref{eq:1}) is an
identity about determinants, whereas what is really needed is an
identity about Pfaffians.  Tracy and Widom get around this difficulty
by using the important observation that, if $\mathbf A$ is an antisymmetric
square matrix of even size, then $\det \mathbf A = (\Pf \mathbf
A)^2$.  

The extra circumlocutions
necessary in the derivation of the correlation functions can be
eliminated by using, in place of (\ref{eq:1}), the fact that
\begin{equation}
\label{eq:2}
\frac{\Pf( \mathbf{C}^{-\transpose} - \mathbf{A}^{\transpose}
  \mathbf{B A})}{\Pf \mathbf C^{-\transpose}} = \frac{\Pf(
  \mathbf{B}^{-\transpose} - \mathbf{A C
    A^{\transpose}})}{\Pf \mathbf B^{-\transpose}},
\end{equation}
where $\mathbf B$ and $\mathbf C$ are arbitrary antisymmetric $2T
\times 2T$ and $2N \times 2N$ matrices with nonzero Pfaffians, and
$\mathbf{A}$ is an arbitrary $2N \times 2T$ matrix.  The derivation of
the correlation functions using (\ref{eq:2}) in the case where $N$ is
even was carried out by Borodin and Sinclair in \cite[Appendix
A]{borodin-2008}.  Equation~(\ref{eq:2}) is Rains' 
Pfaffian Cauchy-Binet formula \cite{rains-2000}; a proof is given in
\cite[Appendix B]{borodin-2008}.   

The main purpose of this note is to show how (\ref{eq:2}) can be used
to derive the Pfaffian form of the correlation functions in the case
when $\beta=1$ and $N$ is odd, thus completing the work started by
Tracy and Widom.  We will also show how (\ref{eq:2}) can be used to
derive the correlation functions of real asymmetric ensembles in the
case when $N$ is odd.  For $N$ even a derivation of the correlation
functions of real asymmetric ensembles using (\ref{eq:2}) was given by
Borodin and Sinclair \cite{borodin-2007}. The work here will
complement the existing methods for deriving the correlation functions
of real asymmetric ensemble given by Sommers and Wieczorek
\cite{sommers-wieczorek} and Mays and Forrester
\cite{mays-forrester}. 

\subsection{Real Asymmetric Matrices}

Ginibre's real ensemble of $N \times N$ matrices is given by $\R^{N
  \times N}$ together with a probability measure specified by treating
the entries of the matrices as independent standard normal random
variables \cite{MR0173726}.  This ensemble is complicated by the fact
that there are two species of eigenvalues: real and complex
conjugate pairs.  Among the implications of this is that there is no one
JPDF for the ensemble.  Instead, we have several partial JPDFs indexed
by pairs non-negative integers $L$ and $M$ satisfying $L + 2M = N$;
$L$ represents the number of real eigenvalues in a `sector' and $M$
the number of pairs of complex conjugate eigenvalues.  These partial
JPDFs are given by
\begin{equation}
\label{eq:14}
\Omega_{L,M}(\bs \alpha, \bs \beta) = \frac{2^M}{Z} \bigg\{ \prod_{\ell=1}^L
w(\alpha_{\ell}) \prod_{m=1}^M w(\beta_m) w(\overline \beta_m) \bigg\}
| \Delta(\bs{\alpha} \vee \bs{\beta} \curlyvee \overline{\bs{\beta}} ) |,
\end{equation}
where $Z$ is a normalizing constant (that depends on $N$ but not $L$
and $M$), $\Delta(\bs \alpha \vee \bs \beta \curlyvee \overline{\bs{\beta}})$ is
the $N \times N$ Vandermonde determinant in the variables 
\[
\alpha_1, \ldots, \alpha_L, \beta_1, \overline{\beta}_1, \ldots,
\beta_M, \overline{\beta}_M,
\]
and $w: \C \rightarrow [0, \infty)$ is a weight function given by
\[
w(\lambda) = e^{-\lambda^2/2} \sqrt{\erfc(\sqrt{2} | \ip \lambda |)}.
\]
When $L=N$ and $M=0$, (\ref{eq:14}) was established by Ginibre in his founding
treatise on real asymmetric ensembles \cite{MR0173726}.  The general
case was established three decades later independently by Lehmann and
Sommers \cite{1991PhRvL..67..941L} and Edelman\cite{MR1437734}.  

We may generalize Ginibre's real ensemble to other ensembles of real
asymmetric matrices by allowing $w$ to be another function, and we
call such an ensemble the real asymmetric ensemble with weight
function $w$; here we will only consider weight functions satisfying
$w(\lambda) = w(\overline \lambda)$ for all $\lambda \in \C$.  

Given non-negative integers $\ell$ and $m$ with $\ell + 2m \leq N$, we
define the $\ell,m$-correlation function of the real asymmetric
ensemble with weight function $w$ by $R_{\ell, m}: \R^{\ell} \times
\C^{m} \rightarrow [0, \infty)$, where 
\begin{align*}
R_{\ell, m}(\mathbf x, \mathbf z) &= \sum_{L \geq \ell, M \geq m}
\frac{1}{(L-\ell)! (M-m)! 2^{M-m}} \\
& \hspace{3cm} \times \int_{\R^{L-\ell}} \int_{\C^{M-m}}
\Omega_{L,M}(\mathbf x \vee \bs \alpha, \mathbf z \vee \bs \beta)
\, d\mu_1^{L-\ell}(\bs \alpha) \, d\mu_2^{M-m}(\bs \beta),
\end{align*}
where $\mu_1$ is Lebesgue measure on $\R$ and $\mu_2$ is Lebesgue
measure on $\C$.  

One goal of the current manuscript is to show that, when $N$ is odd, $R_{\ell,
  m}(\mathbf x, \mathbf z)$ can be written as 
\[
R_{\ell,m}(\mathbf x, \mathbf z) = \Pf \begin{bmatrix}
K_N(x_{j}, x_{j'}) & K_N(x_{j}, z_{k'}) \\
K_N(z_{k}, x_{j'}) & K_N(z_{k}, z_{k'})
\end{bmatrix}_{j,j'=1,\ldots,\ell \atop k,k'=1,\ldots,m},
\]
where $K_N$ is a particular $2 \times 2$ matrix kernel acting on $L^2(\mu_1 +
\mu_2) \times L^2(\mu_1 + \mu_2)$, the exact nature of which will be
explained in the sequel.  A similar statement is true when $N$ is even
\cite{borodin-2007, borodin-2008, sommers-2007, forrester-2007}, and
there are other existing (and arguably more complicated) methods for
the $N$ odd case \cite{sommers-wieczorek, mays-forrester}.

\section{de Bruijn's Identities}

\subsection{For Hermitian Ensemble}

Given a measure $\nu$ on $\R$, we define 
\[
Z^{\nu}_{N,\beta} = \int_{\R^N} | \Delta(\bs \lambda) |^{\beta}
d\nu^N(\bs{\lambda}).
\]
When $d \nu = w d \mu$ for a function $w: \R \rightarrow [0, \infty)$,
$Z_{N,\beta}^{\nu}$ is the normalizing constant for the corresponding ensemble
with weight function $w$.  We call $Z_{N,\beta}^{\nu}$ the {\em
  partition function} of the corresponding ensemble.  

A first step in the derivation of Pfaffian or determinantal form of
the correlation functions for such ensembles is to write
$Z^{\nu}_{N,\beta}$ as a determinant or a Pfaffian.  This can only be
done when $\beta=1,2$ or $4$.  These identities, the first due to
Andr\'eief \cite{andreief} and the second two to de~Bruijn
\cite{MR0079647}, can be 
formulated in our setting as follows: Suppose $p_0(\lambda), p_1(\lambda),
\ldots$ are arbitrary monic polynomials such that $\deg p_n = n$, then
\[
Z^{\nu}_{N, 2} = N! \det \left[\int_{\R} p_j(\lambda) p_k(\lambda)
  d\nu(\lambda) \right]_{j,k=0}^{N-1},
\]
\[
Z^{\nu}_{N,4} = (2N)! \Pf \left[\int_{\R} p_j(\lambda) p_k'(\lambda) -
  p_k(\lambda) p_j'(\lambda) \, d\nu(\lambda) \right]_{j,k=0}^{2N-1}
\]
and when $N$ is even
\[
Z^{\nu}_{N, 1} = \Pf \left[\int_{\R} \int_{\R} p_j(\lambda) p_k(\eta) \,
  \sgn(\eta - \lambda) \, d\nu(\eta) d\nu(\lambda) \right]_{j,k=0}^{N-1}.
\]

We define the $N \times N$ matrix $\mathbf{U}_{N,1}^{\nu}$ by
\begin{equation}
\label{eq:4}
\mathbf{U}_{N,1}^{\nu} = \left[\int_{\R} \int_{\R} p_j(\lambda) p_k(\eta) \,
  \sgn(\eta - \lambda) \, d\nu(\eta) d\nu(\lambda) \right]_{j,k=0}^{N-1}.
\end{equation}
Thus, when $N$ is even, $Z^{\nu}_{N,1} = \Pf \mathbf{U}_{N,1}^{\nu}$.
One reason de~Bruijn did not provide an identity for $Z_{N,
  1}^{\nu}$ for odd $N$ is that the Pfaffian is only defined for
antisymmetric 
square matrices with an even number of rows and columns.  We may
produce a de~Bruijn identity in the odd $N$ case by suitably altering
the matrix to appear on the right hand side of the expression for
$Z_{N ,1}^{\nu}$.  Specifically, for $N$ odd, we have $Z_{N, 1}^{\nu} = \Pf
\mathbf{W}_{N,1}^{\nu}$, where  
\begin{equation}
\label{eq:3}
\mathbf{W}_{N,1}^{\nu} = \begin{bmatrix}
 &  {\displaystyle \int_{\R} p_0(\lambda) d\nu(\lambda)} \\
 {\displaystyle \mathbf{U}_{N,1}^{\nu}} &  \vdots   \\
& {\displaystyle \int_{\R} p_{N-1}(\lambda) d\nu(\lambda)} \\
{\displaystyle -\int_{\R} p_0(\lambda) \, d\nu(\lambda)  \qquad \cdots
  \qquad  -\int_{\R} p_{N-1}(\lambda) \, d\nu(\lambda)} &  0
\end{bmatrix}.
\end{equation}

\subsection{for Real Asymmetric Ensembles}

The corresponding identity for real asymmetric ensembles is given
in \cite{sinclair-2007}.  This can be written down as follows:
Given a measure $\nu_1$ on $\R$ and a measure $\nu_2$ on $\C \setminus
\R$ which is invariant under complex conjugation, define $\nu = \nu_1
+ \nu_2$ and set 
\[
Z_N^{\nu} = \sum_{(L,M) \atop L + 2M = N} \frac{1}{2^M M! L!} \int_{\R^L}
\int_{\C^M} | \Delta(\bs \alpha \vee \bs \beta \curlyvee \overline{\bs
  \beta}) | \, d\nu_1^L(\bs \alpha) d\nu_2^M(\bs \beta).
\]
If $d\nu_1 = w d\mu_1$ and $d\nu_2(\beta) = |w|^2
d\mu_2$ for some function $w: \C \rightarrow [0,\infty)$, then
$Z_N^{\nu}$ is the normalizing constant for the real asymmetric
ensemble with weight function $w$.  When $N$ is even,
\[
Z_{N}^{\nu} = \Pf \mathbf{U}^{\nu}_{N},
\]
where $\mathbf{U}^{\nu}_{N}$ is the $N \times N$ antisymmetric
matrix given by 
\[
\mathbf U_{N}^{\nu} = \left[\; \int\limits_{\R^2} p_j(\alpha) p_k(\gamma) 
  \sgn(\gamma - \alpha) \, d\nu_1(\gamma) d\nu_1(\alpha) -2i \int\limits_{\C}
  p_j(\beta) p_k(\overline \beta) \sgn \ip \beta \, d\nu_2(\beta)
\right]_{j,k=0}^{N-1}
\]
and when $N$ is odd, $Z_N^{\nu} = \Pf \mathbf W_{N}^{\nu}$ where
\[
\mathbf{W}_N^{\nu} = \begin{bmatrix}
 &  {\displaystyle \int_{\R} p_0(\lambda) d\nu_1(\lambda)} \\ \mathbf
 U_{N}^{\nu}&  \vdots   \\ 
& {\displaystyle \int_{\R} p_{N-1}(\lambda) d\nu_1(\lambda)} \\
{\displaystyle -\int_{\R} p_0(\lambda) \, d\nu_1(\lambda)  \qquad \cdots
  \qquad  -\int_{\R} p_{N-1}(\lambda) \, d\nu_1(\lambda)} &  0
\end{bmatrix}.
\]

\section{Correlation Functions in terms of Partition Functions}

From here forward we will limit our attention to $\beta=1$ Hermitian
ensembles and real asymmetric ensembles.  We will drop the subscripts
on all relevant quantities for these ensembles so that, for instance,
$Z^{\nu}$ represents both $Z^{\nu}_{N,1}$ and $Z^{\nu}_{N}$; which
partition function is being represented will be clear from context.
This will allow us to treat these cases simultaneously at the most
technical part of the proof.  

We will ultimately be interested in the case when $N$ is odd, but the
results in this section are equally valid for $N$ even. 

\subsection{For Hermitian Ensembles}
We first consider the case of $w1$E. Let $T > N$ be an even integer,
$y_1, y_2, \ldots, y_T \in \R$ and suppose $c_1, c_2, \ldots, c_T$ are
indeterminants.  We define 
the measure  
\[
d \eta(\lambda) = \sum_{t=1}^T c_t w(y_t) d\delta(\lambda - y_t),
\]
where $\delta$ is the probability measure with unit mass at
$x=0$.  We also define the measure $\nu$ by $d \nu = w d\mu$.   
$Z^{\eta + \nu}/Z^{\nu}$ is the generating function for the
correlation functions of $w1$E; this is the content of the following lemma.
\begin{lemma}
\label{lemma:1}
Given an integer $n > 0$ we define $\ul n = \{1,2,\ldots, n\}$.  
\[
\frac{Z^{\eta + \nu}}{Z^{\nu}} = 1 + \sum_{n=1}^T \sum_{\mf t: \ul n
  \nearrow \ul T} \bigg\{\prod_{j=1}^n c_{\mf{t}(j)} \bigg\}
R_n(y_{\mf t(1)}, \ldots, y_{\mf t(n)}).
\]
\end{lemma}
The proof of this lemma follows easily from the definitions of
Α$Z^{\eta + \nu}$ and $R_n$; details can be found in
\cite{borodin-2007}.  

\subsection{For Real Asymmetric Ensemble}

For the real asymmetric ensembles with weight function $w$ we suppose
$U$ and $V$ are even integers greater than $N$ and set 
\[
d \eta(\lambda) = \sum_{u=1}^U a_{u} w(x_{u}) d\delta(\zeta -
x_u) + \sum_{v=1}^V b_v \big( w(z_v) d\delta(\zeta - z_v) + w(\overline
z_v) \delta(\zeta - \overline z_v) \big),
\]
where $x_1, \ldots, x_U \in \R$, $z_1, \ldots, z_V \in \C \setminus
\R$ and $a_1, \ldots, a_U, b_1, \ldots, b_V$ are indeterminants.  If
$\nu = \nu_1 + \nu_2$ is the measure with $d \nu_1 = w d\mu_1$ and
$d\nu_2 = |w|^2 d\mu_2$, then $Z^{\eta + \nu}/Z^{\nu}$
generates the correlation functions of the corresponding real 
asymmetric ensemble.  
\begin{lemma}
\label{lemma:2}
\[
\frac{Z^{\eta + \nu}}{Z^{\nu}} = \sum_{(\ell, m) \atop \ell + 2m \leq 
  N}  \sum_{\mf u: \ul \ell \nearrow \ul U} \sum_{\mf v: \ul m
  \nearrow \ul V}  \bigg\{ \prod_{j=1}^{\ell} a_{\mf u(\ell)}
\prod_{k=1}^m b_{\mf v(k)} \bigg\} R_{\ell, m}(x_{\mf u(1)}, \ldots,
x_{\mf u(\ell)}, z_{\mf v(1)}, \ldots, z_{\mf v(m)} ),
\]
where by convention we will take 
\[
\sum_{\mf u: \ul 0 \nearrow \ul N} \prod_{j=1}^0 a_{\mf u(j)} =
\sum_{\mf v: \ul 0 \nearrow \ul N} \prod_{k=1}^0 b_{\mf v(j)} = 1.
\]
\end{lemma}
We remark that Lemma~\ref{lemma:1} follows from Lemma~\ref{lemma:2} by
setting $b_1 = \cdots = b_V = 0$.  The proof of Lemma~\ref{lemma:2} is
found in \cite{borodin-2007} and follows directly from the definitions
of $Z^{\eta + \nu}$ and $R_{\ell, m}$.  

\section{Using de Bruijn's Identities}  
From here forward we will assume that $N$ is odd.

\subsection{For Hermitian Ensembles} 

Given a measure $\kappa$ on $\R$ we define the operator $\epsilon^{\kappa}:
L^2(\kappa) \rightarrow L^2(\kappa)$ by
\[
\epsilon^{\kappa} f(\lambda) = \frac{1}{2} \int_{\R} f(\xi) \sgn(\lambda
- \xi) \, d\kappa(\lambda).
\]
Using this we define a skew-symmetric bilinear form on $L^2(\kappa)$
given by 
\[
\la f | g \ra^{\kappa} = \int_{\R} (f \epsilon^{\kappa} g - g \epsilon^{\kappa}
f) \, d\kappa.
\]

Notice that $\mathbf{U}^{\nu}$, as given in (\ref{eq:4}), can also be
written as
\[
\mathbf{U}^{\nu} = \left[
\la p_j | p_k \ra^{\nu} 
\right]_{j,k=0}^{N-1},
\]
where, as before $p_0, p_1, \ldots$ is a sequence of monic polynomials
with $\deg p_n = n$.  

Since $Z^{\eta + \nu}/Z^{\nu}$ generates the correlation functions, we turn now
to the investigation of the entries of $\mathbf{W}^{\eta + \nu}$.  It
will be convenient to set $\wt p_n = w p_n; n=1,2,\ldots$, and to
write $\la f | g \ra = \la f | g \ra^{\mu}$ and $\epsilon =
\epsilon^{\mu}$ (recall $\mu$ is Lebesgue measure on $\R$).  
\begin{align}
  \la p_j | p_k \ra_{\eta + \nu} &= \int_{\R^2} p_j(\lambda) p_k(\xi)
  \sgn(\xi - \lambda) d(\nu + \eta)(\lambda) d(\nu +
  \eta)(\xi) \nonumber  \\
  &= \la \wt p_j | \wt p_k \ra_{\mu} + 2 \sum_{t=1}^T c_t \big( \wt
  p_j(y_t) \epsilon \wt p_k(y_t) - \wt p_k(y_t) \epsilon \wt p_j(y_t)
  \big) \nonumber \\
& \hspace{4cm} - \sum_{t=1}^T \sum_{m=1}^T c_t c_m \wt p_j(y_t) \wt
  p_k(y_m) \sgn(y_m - y_t).\label{eq:5}
\end{align}
The other entries in $\mathbf{W}^{\eta + \nu}$ are of the form
\begin{equation*}
\int_{\R} p_j(\lambda) d(\eta + \nu)(\lambda) = \int_{\R} \wt
p_j(\lambda) d\mu(\lambda) + \sum_{t=1}^T c_t \wt p_j(y_t) 
\end{equation*}

\subsection{For Real Asymmetric Ensemble}

First we extend the definition of the $\sgn$ function to $\C$ by
specifying that $\sgn(z) = 0$ if $z \not\in \R$.  Next, given a
measure $\kappa = \kappa_1 + \kappa_2$ on $\C$ we define the operator
$\epsilon^{\kappa}: L^2(\kappa) \rightarrow L^2(\kappa)$ by   
\[
\epsilon^{\kappa} f(\lambda) = \left\{
\begin{array}{ll}
{\displaystyle \frac{1}{2} \int_{\R} f(\xi) \sgn(\lambda - \xi) \,
  d\kappa_1(\lambda) } &
\quad \lambda \in \R; \\ & \\
-i \sgn g(\overline \lambda) \sgn \ip{\lambda} & \quad \lambda \in \C
\setminus \R.
\end{array}
\right.
\]
As in the Hermitian case, we define a skew-symmetric bilinear form
on $L^2(\kappa)$ by
\[
\la f | g \ra^{\kappa} = \int_{\C} (f \epsilon^{\kappa} g - g \epsilon^{\kappa} f )
\, d\kappa.
\]
As before we set $\epsilon$ and $\la f | g \ra$ for $\epsilon^{\mu}$
and $\la f | g \ra^{\mu}$, where in the context of real asymmetric
ensembles, $\mu = \mu_1 + \mu_2$.  

A superficial change of notation will bring the entries of
$\mathbf{W}^{\eta + \nu}$ into a form identical to the Hermitian
case.  We set $T = U + V$, rename the indeterminants $a_1,
\ldots, a_U, b_1, \ldots, b_V$ to $c_1, c_2, \ldots, c_T$ and rename
$x_1, \ldots, x_U, z_1, \ldots, z_V$ to $y_1, y_2, \ldots, y_T$.  If
we also 
define 
\[
d\widehat \delta(\lambda - \gamma) = \left\{
\begin{array}{ll}
d\delta(\lambda - \gamma) & \quad \gamma \in \R; \\ & \\
d\delta(\lambda - \gamma) + d\delta(\lambda - \overline \gamma) & \quad
\gamma \in \C \setminus \R.
\end{array}
\right.
\]
It follows that we may write 
\[
d \eta(\lambda) = \sum_{t=1}^T c_t w(x_t) d\widehat \delta(\lambda - y_t),
\]
and, as in the Hermitian ensemble case,
\begin{align}
  \la p_j | p_k \ra_{\eta + \nu} 
  &= \la \wt p_j | \wt p_k \ra_{\mu} + 2 \sum_{t=1}^T c_t \big( \wt
  p_j(y_t) \epsilon \wt p_k(y_t) - \wt p_k(y_t) \epsilon \wt p_j(y_t)
  \big) \nonumber \\
& \hspace{4cm} - \sum_{t=1}^T \sum_{m=1}^T c_t c_m \wt p_j(y_t) \wt
  p_k(y_m) \sgn(y_m - y_t), \label{eq:6}
\end{align}
and
\[
\int_{\R} p_j(\lambda) d(\eta + \nu)(\lambda) = \int_{\R} \wt
p_j(\lambda) d\mu(\lambda) + \sum_{t=1}^T c_t \wt p_j(y_t).
\]

\section{The Method of Tracy and Widom}

Since we have written $Z^{\eta + \nu}/Z^{\nu}$ in a uniform manner for
both Hermitian ensembles and real asymmetric ensembles we may, at
least for the moment, analyze both cases simultaneously. 

We define $\mathbf{A}$ to be the $N + 1 \times 2T$ matrix given by 
\[
\mathbf{A} = \begin{bmatrix}
\sqrt{2 c_1} \wt p_0(y_1) & \sqrt{2 c_1} \epsilon \wt p_0(y_1) &
& \sqrt{2 c_T} \wt p_0(y_T) & \sqrt{2 c_T} \epsilon \wt p_0(y_T)
\\
\sqrt{2 c_1} \wt p_1(y_1) & \sqrt{2 c_1} \epsilon \wt p_1(y_1) & \cdots
& \sqrt{2 c_T} \wt p_1(y_T) & \sqrt{2 c_T} \epsilon \wt p_1(y_T)
\\
&  \vdots & \ddots & \vdots & \\
\sqrt{2 c_1} \wt p_{N-1}(y_1) & \sqrt{2 c_1} \epsilon \wt p_{N-1}(y_1) & \cdots
& \sqrt{2 c_T} \wt p_{N-1}(y_T) & \sqrt{2 c_T} \epsilon \wt p_{N-1}(y_T)
\\
0 & \chi_{\R}(y_1) \sqrt{c_1/2} & & 0 & \chi_{\R}(y_T) \sqrt{c_T/2}
\end{bmatrix},
\]
where $\chi_{\R}$ is the characteristic function of $\R$. 
That is,
\begin{align*}
\mathbf{A}_{n, 2t-1} = \sqrt{2 c_t} \wt p_n(y_t), & \qquad
n=0,1,\ldots,N-1 \qquad t=1,2,\ldots,T; \\
\mathbf{A}_{n,2t} = \sqrt{2 c_t} \epsilon \wt p_n(y_t), & \qquad
n=0,1,\ldots,N-1 \qquad t=1,2,\ldots,T,
\end{align*}
and
\begin{align*}
&\mathbf{A}_{N,2t-1} = 0, \hspace{2.1cm} t=1,2,\ldots,T; \\
&\mathbf{A}_{N,2t} = \chi_{\R}(y_t) \sqrt{c_t/2}, \qquad t=1,2,\ldots,T.
\end{align*}

and set $\mathbf{J}$ to be the $2T \times 2T$ matrix,
\[
\mathbf{J} = \begin{bmatrix}
0  & 1 &        &    &   \\
-1 & 0 &        &    &   \\
   &   & \ddots &    &   \\
   &   &        & 0  & 1 \\
   &   &        & -1 & 0
\end{bmatrix}.
\]
It follows that,
\[
\mathbf{A J A}^{\transpose} = \begin{bmatrix}
 & \displaystyle{\sum_{t=1}^T c_t \chi_{\R}(y_t) \wt p_0(y_t)} \\
\displaystyle{\left[
2 \sum_{t=1}^T c_t \big( \wt p_n(y_t) \epsilon
  \wt p_m(y_t) - \epsilon \wt p_n(y_t) \wt p_m(y_t) \big)
\right]_{n,m=0}^{N-1}} & \vdots \\ 
                 & \displaystyle{\sum_{t=1}^T c_t \chi_{\R}(y_t) \wt
                   p_{N-1}(y_t)} \\ 
\displaystyle{-\sum_{t=1}^T c_t \chi_{\R}(y_t) \wt p_0(y_t)
  \hspace{.5cm} \cdots \hspace{.5cm}  -\sum_{t=1}^T c_t
  \chi_{\R}(y_t) \wt p_{N-1}(y_t)} & 0 
\end{bmatrix}.
\]

Next, we define the $2T \times 2T$ matrix
\[
\mathbf{E} = \begin{bmatrix}
{\displaystyle \frac{\sqrt{c_t c_u}}{2} \sgn(y_u - y_t)} & 0 \\
0 & 0
\end{bmatrix}_{t,u=1;}^T
\]
so that,
\[
\mathbf{E}_{2t-1,2u-1} = \frac{\sqrt{c_t c_u}}{2} \sgn(y_u - y_t); \qquad
t,u=1,2,\ldots,T,
\]
and all other entries of $\mathbf{E}$ are equal to 0.  An easy
computation reveals
\[
(\mathbf{AEA}^{\transpose})_{n,m} = \sum_{u=1}^T \sum_{t=1}^T c_u
c_t \wt p_m(y_u) \wt p_n(y_t) \sgn(y_u - y_t); \qquad n,m=0,1,\ldots,N-1,
\]
and all entries in the last row and column of
$(\mathbf{AEA}^{\transpose})$ are 0.  That is,
\[
\mathbf{A E A}^{\transpose} = \begin{bmatrix}
 & 0 \\
\displaystyle{\left[
2 \sum_{u=1}^T \sum_{t=1}^T c_u c_t \wt p_m(y_u)  \wt p_n(y_t)
\mathcal{E}(y_t, y_u) 
\right]_{n,m=0}^{N-1}} & \vdots \\ 
                 & 0 \\ 
0
  \hspace{2cm} \cdots \hspace{2cm}  0 & 0 
\end{bmatrix}.
\]

It follows that
\[
\mathbf{W}^{\eta + \nu} = \mathbf{W}^{\nu} +
\mathbf{A J A}^{\transpose} - \mathbf{A E A}^{\transpose}.
\]
And, if we define the $N +1 \times N+1$ matrices
$\mathbf{C}^{-\transpose} = \mathbf{W}^{\nu}$ and
$\mathbf{B} = -\mathbf{J} + \mathbf{E}$, then 
\begin{equation}
\label{eq:10}
\frac{Z^{\nu + \eta}}{Z^{\nu}} = \frac{\Pf \mathbf{W}^{\nu
    + \eta} }{\Pf \mathbf{W}^{\nu} } = \frac{\Pf \big(
  \mathbf{C}^{-\transpose} - \mathbf{A B A}^{\transpose} \big)}{\Pf
\mathbf{C}^{-\transpose}}.
\end{equation}
We are now in position to use the identity
\begin{equation}
\label{eq:8}
\frac{\Pf \big(
  \mathbf{C}^{-\transpose} - \mathbf{A B A}^{\transpose} \big)}{\Pf
\mathbf{C}^{-\transpose}} = \frac{\Pf \left(
  \mathbf{B}^{-\transpose} - \mathbf{A}^{\transpose}\mathbf{C A} \right)}{\Pf
\mathbf{B}^{-\transpose}},
\end{equation}
and an easy calculation shows that
\[
\mathbf{B}^{-\transpose} = - \mathbf{J} - \mathbf{E}',
\]
where $\mathbf{E}'$ is the $2T \times 2T$ matrix,
\[
\mathbf{E}' = \begin{bmatrix}
0 & 0 \\
0 & {\displaystyle \frac{\sqrt{c_t c_u}}{2} \sgn(y_u - y_t) }
\end{bmatrix}_{t,u=1}^T.
\]
It is also easily verified that $\Pf \mathbf{B}^{-\transpose} =
(-1)^T$. Thus,
\[
\frac{Z^{\nu + \eta}}{Z^{\nu}} = (-1)^T  \Pf\big(-\mathbf{J} -
\mathbf{E}' - \mathbf{A}^{\transpose} \mathbf{C A} \big) = \Pf\big(\mathbf{J} +
\mathbf{E}' + \mathbf{A}^{\transpose} \mathbf{C A} \big).
\]

We now compute the entries of $\mathbf{A}^{\transpose}\mathbf{CA}$:
\begin{align*}
& (\mathbf{A}^{\transpose}\mathbf{CA})_{2u-1, 2t-1} = 2 \sqrt{c_u c_t}
\sum_{n=0}^{N-1}
\sum_{m=0}^{N-1}  \wt p_n(y_u) \mathbf{C}_{n,m} \wt p_m(y_t); \qquad
u,t=1,2,\ldots, T; \\
& (\mathbf{A}^{\transpose}\mathbf{CA})_{2u-1, 2t} = 2 \sqrt{c_u c_t} \sum_{n=0}^{N-1}
\sum_{m=0}^{N-1}  \wt p_n(y_u) \mathbf{C}_{n,m} \epsilon \wt
p_m(y_t) + \sqrt{c_u c_t} \chi_{\R}(y_t) \sum_{n=0}^{N-1} \wt p_n(y_u) \mathbf{C}_{n,N}
; \\
& \hspace{9.3cm} u,t=1,2,\ldots,T; \\
 \\
& (\mathbf{A}^{\transpose}\mathbf{CA})_{2u, 2t-1} = 2 \sqrt{c_u c_t} \sum_{n=0}^{N-1}
\sum_{m=0}^{N-1}  \epsilon \wt p_n(y_u) \mathbf{C}_{n,m} \wt
p_m(y_t) + \sqrt{c_u c_t} \chi_{\R}(y_u) \sum_{m=0}^{N-1} \mathbf{C}_{N,m} \wt p_m(y_t); \\
& \hspace{9.3cm} u,t=1,2,\ldots,T; \\
& (\mathbf{A}^{\transpose}\mathbf{CA})_{2u, 2t} = \sqrt{c_u c_t}
\chi_{\R}(y_t) \sum_{n=0}^{N-1}  
\mathbf{C}_{n,N} \epsilon \wt p_n(y_u) +  \sqrt{c_u
  c_t} \chi_{\R}(y_u) \sum_{m=0}^{N-1}  \mathbf{C}_{N,m} \epsilon \wt p_m(y_t); \\
& \hspace{2.2cm} + 2 \sqrt{c_u c_t} \sum_{n=0}^{N-1}
\sum_{m=0}^{N-1}  \epsilon \wt p_n(y_u) \mathbf{C}_{n,m} 
\epsilon \wt p_m(y_t)  ; 
 \qquad u,t=1,2,\ldots,T.
\end{align*}
If we define
\begin{align*}
DS_N(y,y') &= 2 \sum_{n=0}^{N-1}
\sum_{m=0}^{N-1}  \wt p_n(y) \mathbf{C}_{n,m} \wt p_m(y'); \\ 
S_N(y,y') &= 2 \sum_{n=0}^{N-1}
\sum_{m=0}^{N-1}  \wt p_n(y) \mathbf{C}_{n,m} \epsilon \wt
p_m(y') +  \chi_{\R}(y') \sum_{n=0}^{N-1} \wt p_n(y) \mathbf{C}_{n,N}
; \\
S_NI(y,y') &= 2 \sum_{n=0}^{N-1} \sum_{m=0}^{N-1}  \epsilon \wt p_n(y)
\mathbf{C}_{n,m} \epsilon \wt p_m(y') \\ 
& \hspace{3cm} + \sum_{n=0}^{N-1}
\mathbf{C}_{n,N} \big( \chi_{\R}(y') \epsilon \wt p_n(y) -
\chi_{\R}(y) \epsilon \wt p_n(y') \big),
\end{align*}
then
\begin{equation}
\label{eq:9}
\mathbf{A}^{\transpose} \mathbf{CA} = \Bigg[
\sqrt{c_u c_t} \begin{bmatrix}
DS_N(y_u, y_t) & S_N(y_u, y_t) \\
-S_N(y_t, y_u) & S_NI(y_u,y_t) 
\end{bmatrix}
\Bigg]_{u,t=1}^T,
\end{equation}
and, if we define the $2 \times 2$ matrix kernel $K_N: \C
\times \C \rightarrow \C^{2 \times 2}$ by
\[
K_N(y,y') = \begin{bmatrix}
DS_N(y, y') & S_N(y, y') \\
-S_N(y, y') & S_NI(y,y') +  \frac{1}{2}\sgn(y' - y) 
\end{bmatrix},
\]
and the $2T \times 2T$ matrix $\mathbf{K}$ by
\[
\mathbf{K} = \big[ \sqrt{c_u c_t} K_N(y_u, y_t) \big]_{u,t=1}^T
\]
then from (\ref{eq:10}), (\ref{eq:8}), (\ref{eq:9}) we see that
\begin{equation}
\label{eq:12}
\frac{Z^{\nu + \eta}}{Z^{\nu}} = \Pf \left( \mathbf{J} + \mathbf{K} \right).
\end{equation}

\section{Recovering the Correlation Functions}

The final step in the derivation of the correlation functions is to
expand $\Pf(\mathbf{J} + \mathbf{K})$ using the identity
\begin{equation}
\label{eq:7}
\Pf( \mathbf J + \mathbf K) = 1 + \sum_{n=1}^{T} \sum_{\mf t: \ul{n}
  \nearrow \ul T} \Pf \mathbf{K}_{\mf t},
\end{equation}
where $\mathbf{K}_{\mf t}$ is the $2n \times 2n$ minor of
$\mathbf{K}$ given by
\[
\mathbf{K}_{\mf t} = \left[ \sqrt{c_{\mf t(j)} c_{\mf t(k)}}
  K_N(y_{\mf t(j)}, y_{\mf t(k)})
\right]_{j,k=1}^{n}.
\]
A Proof of this identity can be found in \cite{MR1069389} or
\cite{sinclair-2005}.  

\subsection{For Hermitian Ensembles}

Applying (\ref{eq:7}) to (\ref{eq:12}) and using the fact that
\[
\Pf \mathbf{K}_{\mf t, \mf t} = \bigg\{ \prod_{m=1}^n c_{\mf t(m)}
\bigg\} \Pf \left[ K_N(y_{\mf t(j)}, y_{\mf{t}(k)}) \right]_{j,k=1}^n,
\]
we find that 
\[
\frac{Z^{\nu + \eta}}{Z^{\nu}} = 1 + \sum_{n=1}^T \sum_{\mf t: \ul n
  \nearrow \ul T} \bigg\{ \prod_{m=1}^n c_{\mf t(m)}
\bigg\} \Pf \left[ K_N(y_{\mf t(j)}, y_{\mf{t}(k)}) \right]_{j,k=1}^n.
\]
This together with Lemma~\ref{lemma:1} implies that 
\[
R_n(\mathbf y) = \Pf \left[ K_N(y_{j}, y_{k}) \right]_{j,k=1}^n.
\]

\subsection{For Real Asymmetric Ensembles}

Given $\mf u: \ul \ell \nearrow \ul U$ and $\mf v: \ul m \nearrow \ul
V$, we define $\mf u \vee \mf v: \ul{\ell + m} \nearrow \ul{U + V}$ by
\[
(\mf u \vee \mf v)(j) = \left\{
\begin{array}{ll}
\mf u(j) & \quad j \leq \ell \\
\ell + \mf v(j) & \quad j > \ell.
\end{array}
\right.
\]
If $\ell + m = n$, then each $\mf t: \ul n \nearrow \ul T$ can be
written as $\mf u \vee \mf v$ for some $\mf u: \ul \ell \nearrow \ul
U$ and $\mf v : \ul m \nearrow \ul V$.  (This does not preclude the
possibility that $\mf t = \mf u$ or $\mf t = \mf v$).  Using
(\ref{eq:7}) we find
\[
\Pf(\mathbf J + \mathbf K) = \sum_{(\ell, m) \atop \ell \leq U, m \leq V} \sum_{\mf u: \ul \ell \nearrow \ul U} \sum_{\mf v: \ul m
  \nearrow \ul V} \Pf \mathbf{K}_{\mf u \vee \mf v},
\]
where we take the Pfaffian of an empty matrix ({\it i.e.}~when $\ell$
and $m$ equal 0) to be 1.  It is straightforward, if technical, to
show that 
\[
\Pf \mathbf{K}_{\mf u \vee \mf v} = \bigg\{
\prod_{j=1}^{\ell} a_{\mf u(j)} \prod_{k=1}^m b_{\mf v(k)}\bigg\}
\Pf \begin{bmatrix}
K_N(x_{\mf u(j)}, x_{\mf u(j')}) & K_N(x_{\mf u(j)}, z_{\mf v(k')}) \\
K_N(z_{\mf v(k)}, x_{\mf u(k')}) & K_N(z_{\mf v(k)}, z_{\mf v(k')})
\end{bmatrix}_{j,j'=1,\ldots,\ell \atop k,k'=1,\ldots,m}.
\]
Thus, from~(\ref{eq:12}),
\begin{equation}
\label{eq:11}
\frac{Z^{\nu + \eta}}{Z^{\nu}} = \!\!\!\!\!\! \sum_{(\ell, m) \atop 0 \leq \ell + m
  \leq T} \sum_{\mf u: \ul \ell \nearrow \ul U} \sum_{\mf v: \ul m
  \nearrow \ul V} \bigg\{ \prod_{j=1}^{\ell} a_{\mf u(j)}
\prod_{k=1}^m b_{\mf v(k)}\bigg\}
\Pf \begin{bmatrix}
K_N(x_{\mf u(j)}, x_{\mf u(j')}) & K_N(x_{\mf u(j)}, z_{\mf v(k')}) \\
K_N(z_{\mf v(k)}, x_{\mf u(k')}) & K_N(z_{\mf v(k)}, z_{\mf v(k')})
\end{bmatrix}.
\end{equation}
This result and Lemma~\ref{lemma:2} imply that
\[
R_{\ell,m}(\mathbf x, \mathbf z) = \Pf \begin{bmatrix}
K_N(x_{j}, x_{j'}) & K_N(x_{j}, z_{k'}) \\
K_N(z_{k}, x_{j'}) & K_N(z_{k}, z_{k'})
\end{bmatrix}_{j,j'=1,\ldots,\ell \atop k,k'=1,\ldots,m}.
\]

\section{Simplifying the Matrix Kernel}

On first inspection, the kernel $K_N$ is dependent on 
$(p_0, p_2, \ldots, p_{N-1})$.  In fact, 
$K_N$ can be shown to be independent of this family of moni
polynomials ---though a wise choice may simplify the representation of
$K_N$.  In this section we suppose the existence of a complete monic
family of polynomials $\mathbf{q} = ( q_0, q_1, \ldots, q_{N-1})$ such
that 
\[
\la \wt q_{2n} | \wt q_{2m-1} \ra_{\nu} = - \la \wt q_{2m-1} | \wt q_{2n}
\ra_{\nu} = \delta_{n,m} r_n \qquad n=0,1,\ldots,J,
\]
where $J = (N-1)/2-1$.  Here $\mathbf{r} = (r_1, r_2, \ldots,
r_J )$ are referred to the {\em normalizations} of
$\mathbf{q}$.  We also define
\[
s_n = \int_{\R} \wt q_n(\alpha) \, d\nu_{\R}(\alpha), \qquad n=0,1,\ldots,N-1.
\]
Clearly $\mathbf{q}, \mathbf{r}$ and $\mathbf{s}$ are dependent on
$w$.

Using these definitions, $\mathbf W^{\nu}$ becomes 
\[
\mathbf{W}^{\nu} = \begin{bmatrix}
0    & r_0 &      &        &        &      &     & 0 & s_0    \\
-r_0 & 0   &      &        &        &      &     & 0 & s_1    \\
     &     & 0    & r_1    &        &      &     & 0 & s_2    \\
     &     & -r_1 & 0      &        &      &     & 0 & s_3    \\
     &     &      &        & \ddots &      &     & \vdots & \vdots \\
     &     &      &        &        & 0    & r_J & 0 & s_{N-3}    \\
     &     &      &        &        & -r_J & 0   & 0 & s_{N-2}     \\
0    & 0   & 0    & 0      & \cdots      & 0    & 0   & 0 & s_{N-1} \\
-s_0 & -s_1 & -s_2 & -s_3   & \cdots & -s_{N-3} & -s_{N-2} & -s_{N-1} & 0 
\end{bmatrix}.
\]
An easy computation reveals,
\[
Z^{\nu} = \Pf \mathbf{W}^{\nu} = s_{N-1} \prod_{j=1}^J r_j.
\]

In order to present the entries of $K_N$ in their most simplified
form, we need to invert $\mathbf{W}^{\nu}$.  Recall that
$\mathbf C = (\mathbf{W}^{\nu})^{-\transpose}$.
Explicitly, $\mathbf{C}$ is given by
\[
\begin{bmatrix}
0    & \displaystyle{\frac{1}{r_0}} &      &        &        &      &     &
{\displaystyle \frac{-s_1}{r_0 s_{N-1}}} & 0    \\ 
{\displaystyle -\frac{1}{r_0}} & 0   &      &        &        &      &       &
\displaystyle{\frac{s_0}{r_0 s_{N-1}}} & 0    \\
     &     & 0    & {\displaystyle \frac{1}{r_1}}    &        &      &     &
     {\displaystyle \frac{-s_3}{r_1 s_{N-1}}}  & 0    \\ 
     &     & {\displaystyle -\frac{1}{r_1}} & 0      &        &      &       &
     {\displaystyle \frac{s_2}{r_1 s_{N-1}}} & 0    \\ 
     &     &      &        & \ddots &      &                & \vdots & \vdots \\
     &     &      &        &        & 0    & {\displaystyle
      \frac{1}{r_J}} & {\displaystyle \frac{-s_{N-2}}{r_J s_{N-1}}} & 0    \\ 
     &     &      &        &        & {\displaystyle -\frac{1}{r_J}} &
     0     & {\displaystyle \frac{s_{N-3}}{r_J s_{N-1}}} & 0     \\ 
{\displaystyle \frac{s_1}{r_0 s_{N-1}}} & {\displaystyle \frac{-s_0}{r_0
    s_{N-1}}} & {\displaystyle \frac{s_3}{r_1 s_{N-1}}} & {\displaystyle
  \frac{-s_2}{r_1 s_{N-1}}}   & \cdots & 
{\displaystyle \frac{s_{N-2}}{r_J s_{N-1}}} & {\displaystyle
  \frac{-s_{N-3}}{r_J s_{N-1}}}    & 0 & {\displaystyle \frac{1}{s_{N-1}}}  \\  
0    & 0   & 0    & 0      & \cdots      & 0    & 0         &
{\displaystyle \frac{-1}{s_{N-1}}} &
0
\end{bmatrix}.
\]
At the entry level,
\begin{align*}
&\mathbf{C}_{2j, 2m+1} = -\mathbf{C}_{2j+1, 2m} = \frac{\delta_{j,m}
}{r_j}, \qquad \mathbf{C}_{2j,2m} = \mathbf{C}_{2j+1,2m+1} = 0,
 \qquad j,m=0,1,\ldots,J; \\
&\mathbf{C}_{2j, N-1} = -\mathbf{C}_{N-1,2j} = -\frac{s_{2j+1}}{r_j s_{N-1}},
\;\; \mathbf{C}_{2j+1, N-1} = -\mathbf{C}_{N-1,2j+1} = \frac{s_{2j}}{r_j
  s_{N-1}}, \quad j=0,1,\ldots,J;
\end{align*}
and
\[
\mathbf{C}_{N, m} = -\mathbf{C}_{m,N}
= -\frac{\delta_{m+1,N}}{s_{N-1}}, \qquad m=0,1\dots,N.
\]

It follows that 
\begin{align*}
& DS_N(y,y') = 2 \sum_{j=0}^J \frac{\wt q_{2j}(y) \wt q_{2j+1}(y') - \wt
  q_{2j+1}(y) \wt q_{2j}(y')}{r_j} \\
& \quad + \frac{2 \wt q_{N-1}(y)}{s_{N-1}}
\sum_{j=0}^J \frac{s_{2j+1} \wt q_{2j}(y') - s_{2j} \wt
  q_{2j+1}(y')}{r_j}
 + \frac{2 \wt q_{N-1}(y')}{s_{N-1}}
\sum_{j=0}^J \frac{s_{2j} \wt q_{2j+1}(y) - s_{2j+1} \wt
  q_{2j}(y)}{r_j}, \\
& S_N(y,y') = 2 \sum_{j=0}^J \frac{\wt q_{2j}(y) \wt \epsilon
  q_{2j+1}(y') - \wt q_{2j+1}(y) \epsilon \wt q_{2j}(y')}{r_j} \\
& \hspace{4cm} + \frac{2 \wt q_{N-1}(y)}{s_{N-1}}
\sum_{j=0}^J \frac{s_{2j+1} \epsilon \wt q_{2j}(y') - s_{2j}
  \wt \epsilon q_{2j+1}(y')}{r_j} \\
& \hspace{4cm} + \frac{2 \epsilon \wt q_{N-1}(y')}{s_{N-1}}
\sum_{j=0}^J \frac{s_{2j} \wt q_{2j+1}(y) - s_{2j+1} \wt
  q_{2j}(y)}{r_j} + \frac{\wt q_{N-1}(y) \chi_{\R}(y')}{s_{N-1}},
\end{align*}
and
\begin{align*}
& S_NI(y,y') = 2 \sum_{j=0}^J \frac{\epsilon\wt q_{2j}(y) 
  \epsilon \wt q_{2j+1}(y') - \epsilon \wt q_{2j+1}(y)
  \epsilon \wt q_{2j}(y')}{r_j} \\ 
& \hspace{4cm} + \frac{2 \epsilon \wt q_{N-1}(y)}{s_{N-1}}
\sum_{j=0}^J \frac{s_{2j+1} \epsilon \wt q_{2j}(y') - s_{2j}
  \epsilon \wt q_{2j+1}(y')}{r_j} \\
& \hspace{.25cm} + \frac{2 \epsilon \wt q_{N-1}(y')}{s_{N-1}}
\sum_{j=0}^J \frac{s_{2j} \epsilon \wt q_{2j+1}(y) - s_{2j+1}
  \epsilon \wt 
  q_{2j}(y)}{r_j} + \frac{\epsilon \wt q_{N-1}(y) \chi_{\R}(y') -
  \epsilon \wt q_{N-1}(y') \chi_{\R}(y)}{s_{N-1}},
\end{align*}

\bibliography{bibliography}

\begin{center}
\noindent\rule{4cm}{.5pt}
\vspace{.25cm}

\noindent {\sc \small Christopher D.~Sinclair}\\
{\small Department of Mathematics, University of Colorado, Boulder CO 80309} \\
email: {\tt chris.sinclair@colorado.edu}
\end{center}

\end{document}